\documentclass[prb,showpacs,showkeys,preprintnumbers,amsmath,amssymb,twocolumn]{revtex4-1}
\usepackage{graphicx} % Incluir graficos no arquivo.
\usepackage{subfigure}
\usepackage{epstopdf}
\begin{document}

%\title{Thermal entanglement hundreds of Kelvin above room temperature in Iron dimer}
\title{Evidence for entanglement at high temperatures in an engineered molecular magnet}

\author{M. S. Reis and S. Soriano}\affiliation{Instituto de F\'{i}sica, Universidade Federal Fluminense, Av. Gal. Milton Tavares de Souza s/n, 24210-346, Niter\'{o}i-RJ, Brazil}
\author{A. M. dos Santos}
\affiliation{Quantum Condensed Matter Div., Oak Ridge National Lab., Oak Ridge, TN 37831-6475, USA}
\author{B. C. Sales}\affiliation{Materials Science and Engineering Div., Oak Ridge National Lab., Oak Ridge, TN 37831-6056 USA}
\author{D. O. Soares-Pinto}\affiliation{Instituto de F\'{i}sica de S\~{a}o Carlos, Universidade de S\~{a}o Paulo, P.O. Box 369, S\~{a}o Carlos, 13560-970 SP, Brazil}
\author{P. Brand\~{a}o}\affiliation{CICECO, Universidade de Aveiro, 3810-205, Aveiro, Portugal}

\date{\today}

\begin{abstract}
The molecular compound [Fe$_{2}$($\mu_{2}$-oxo)(C$_{3}$H$_{4}$N$_{2}$)$_{6}$(C$_{2}$O$_{4}$)$_{2}$] was designed and synthesized for the first time and its structure determined using single crystal x-ray diffraction. The magnetic susceptibility of this compound was measured from 2 to 300 K.  Analysis of the susceptibility data using protocols developed for other spin singlet ground state systems indicates the existence of entangled quantum states up to temperatures as high as 732 K, the highest entanglement temperature reported to date. The large gap between the ground state and the first excited state (282 K) suggests that the spin system may be somewhat immune to decohering mechanisms. Our measurements strongly suggest that molecular magnets are promising candidate platforms for quantum information processing.  
\end{abstract}

\maketitle

\section{Introduction}

Entanglement, a sort of quantum correlation, plays an important role in Quantum Information Science because it can be used as resource for many quantum information processing protocols\cite{livro_Vedral}. Therefore, a great deal of research aims at determining its existence and strength in a given system. From a theoretical point of view, an entangled state can be defined as a state whose density matrix cannot be written as a convex sum of product states\cite{2009_RMP_81_865}. Following this definition, some measures of entanglement can be constructed, for instance entanglement of formation\cite{1998_PRL_80_2245} or negativity\cite{2002_PRA_65_032314}, and several necessary and sufficient conditions exist that allow the quantification of these quantum correlations\cite{livro_Bruss}.

Nevertheless, the experimental quantification of entanglement is a rather complicated task due to the limited information one can obtain about the system. In some cases, experimentalists need to determine the density matrix of the system, for example using quantum state tomography, and then calculate the appropriate quantity (such as the entanglement of formation or the negativity), which makes the process very demanding\cite{2002_PRL_89_127902,2009_PhysRep_474_1}. An alternative way is not to quantify the degree of entanglement in the system but to estimate it, for example, through the use of an entanglement witness\cite{2011_PRL_107_240502,2010_APhysB_98_617,2010_NJP_12_053007,2007_PRL_99_250405,2004_PRA_69_052327,2012_JPhys_A_45_105302}. Such witnesses are observables with positive expectation value for all separable states and thus a negative expectation value indicates the presence of entanglement\cite{2009_PhysRep_474_1,2009_PRA_79_042334, 2006_PRA_74_010301,2005_PRA_72_022340}. This approach is extremely useful the direct detection the contribution of entangled states to the measurement allows the mapping of these witnesses to the experimental data\cite{nature_425_2003_48,2005_NJP_7_258}.

In many systems, entanglement is a very fragile correlation which can easily vanish due to the coupling between the system and the reservoir, even if this coupling is weak\cite{2009_Science_323_598}. Due to this fact, it was thought that entanglement could only exist in low-temperature few particle systems. However, recently, it has been shown that entanglement can also exist in systems containing a large number of particles at finite temperatures\cite{2008_RMP_80_517,2001_PRL_87_017901,2005_PRB_71_153105,2011_JPhysA_44_025001,2009_EPJB_72_491,2009_PRL_102_100503,2008_ELP_81_40006,2009_PRA_79_052337,2005_PRA_71_010301,2006_APhysB_82_237}. In this sense, the use of entanglement witness has helped to demonstrate, through the measurement of some thermodynamic observables, the existence of entangled states in thermal systems even at high temperature\cite{2006_PRB_73_134404,2007_PRB_75_054422,2008_PRB_77_104402,2009_PRB_79_054408,2009_EPL_87_40008,2008_EPL_84_60003,2006_PRA_74_012314}.

Therefore, the study of entanglement in solid state physics is of great relevance to the area of quantum information, since many proposals of quantum processors are solid state based\cite{2008_RMP_80_517}. The recent demonstration that entanglement can change the thermodynamic properties of solids, such as magnetic susceptibility\cite{nature_425_2003_48,2005_NJP_7_258}, shows that entanglement can be related to significant macroscopic effects. Hence, this subject establishes an interesting connection between quantum information theory and condensed matter physics, because magnetization\cite{arqve}, heat capacity\cite{prb_78_2008_064108}, and internal energy\cite{pla_301_2002_1} can also be used to reveal spin entanglement among constituents of a solid.

Quantum entanglement at elevated temperatures has been studied in several physical systems and the recent work by Vedral\cite{sciam_vedral} summarizes this scenario. In that work it is recognized that some molecular magnets remain entangled at surprisingly high temperatures. Besides presenting robust entangled states, such materials can be engineered to enhance their quantum features and, consequently, to better suited for quantum information applications. As an example, it was  recently shown through magnetic susceptibility measurements, that certain families of molecular magnets are entangled up to 630 K, and have the maximum degree of entanglement at temperatures as high as 100 K\cite{2008_PRB_77_104402,2009_PRB_79_054408,2009_EPL_87_40008}. 

In this paper, we report the synthesis, crystal structure and magnetic susceptibility of a new molecular magnet: [Fe$_{2}$($\mu_{2}$-oxo)(C$_{3}$H$_{4}$N$_{2}$)$_{6}$(C$_{2}$O$_{4}$)$_{2}$]. Analysis of susceptibility data suggests the existence of entangled quantum states up to temperatures as high as 732 K, the highest temperature reported to date.

\section{Thermal entanglement witnessing}

Considering that the total Hamiltonian of a system is given by $\mathcal{H} = \mathcal{H}_{0} + \mathcal{H}_{1}$, where $\mathcal{H}_{0}$ is the spins Hamiltonian and $\mathcal{H}_{1} = g\mu_{B}B\sum_{i=1}^{N} S_{z}^{i}$ is the Zeeman interation of each spin under the constraint $[\mathcal{H}_{0}, \mathcal{H}_{1}] = 0$. As not all spin Hamiltonians commute with the Zeeman term, it is important that the physical realization of a quantum entagled system should allow modeling through an Hamiltonian that does obey the constraint. 

Considering this constraint, the magnetic susceptibility can be written in terms of the correlation functions as:
\begin{eqnarray}
\chi_{\alpha} &=& \frac{(g\mu_{B})^{2}}{k_{B}T}\Delta^{2}M_{\alpha} \nonumber \\
&=& \frac{(g\mu_{B})^{2}}{k_{B}T} \left(\sum_{i,j=1}^{N}\left< S_{i,\alpha}S_{j,\alpha}\right> - \left< \sum_{i=1}^{N} S_{i,\alpha}\right>^{2} \right) \label{susceptib}
\end{eqnarray}
where $g$ is the Land\'{e} factor, $\mu_B$ is the Bohr magneton, and $\Delta^{2}M_{\alpha}$ is the variance of the magnetization in the $\alpha-$direction ($\alpha = x, y, z$).

It must be noted that, for a arbitrary state of a spin$-S$ particle, one has \cite{2003_PRA_68_032103}:
\begin{equation}\label{var1}
\left<S_{x}^{2}\right> + \left<S_{y}^{2}\right> + \left<S_{z}^{2}\right> = S(S+1)
\end{equation}
and
\begin{equation}\label{var2}
\left<S_{x}\right>^{2} + \left<S_{y}\right>^{2} + \left<S_{z}\right>^{2} \leq S^{2}
\end{equation}
The first relation comes from the fact that the eigenvalue of the operator $S^{2}$ is $S(S+1)$, and the second that the projection of the spin in any direction cannot be higher the $S$. Taking the difference between these two equations it is found that:
\begin{equation}\label{var3}
\Delta^{2}S = \Delta^{2}S_{x} + \Delta^{2}S_{y} + \Delta^{2}S_{z} \geq S
\end{equation}

Meanwhile, for any separable state of $N-$spins, the density operator of the system is a convex sum of product
states:
\begin{equation}\label{rhosep}
\rho = \sum_{n} p_{n}\,\rho_{1}^{(n)} \otimes \rho_{2}^{(n)} \otimes\ldots\otimes \rho_{N}^{(n)}
\end{equation}
And if it is substituted into:
\begin{equation}\label{suscepav}
\bar{\chi} = \chi_{x} + \chi_{y} + \chi_{z} = \frac{(g\mu_{B})^{2}}{k_{B}T}\left(\Delta^{2}M_{x} + \Delta^{2}M_{y} + \Delta^{2}M_{z}\right)
\end{equation}
Together with Eq.(\ref{var3}) one find that:
\begin{eqnarray}
\bar{\chi} &=& \frac{(g\mu_{B})^{2}}{k_{B}T}\sum_{n} p_{n} \sum_{i=1}^{N}\left[\left(\Delta^{2}S_{x}\right)_{i}^{(n)} + \left(\Delta^{2}S_{y}\right)_{i}^{(n)} + \left(\Delta^{2}S_{z}\right)_{i}^{(n)}\right] \nonumber \\
&=& \frac{(g\mu_{B})^{2}}{k_{B}T}\sum_{n} p_{n} \Delta^{2}S_{i}^{(n)} \nonumber \\
&\geq& \frac{(g\mu_{B})^{2}N\,S}{k_{B}T} \label{susceptdesig}
\end{eqnarray}

In the case of isotropic systems, where $\chi_{x} = \chi_{y} = \chi_{z}$, this can be reduced to:
\begin{equation}\label{suscepav2}
\bar{\chi} \geq \frac{(g\mu_{B})^{2}N\,S}{3\,k_{B}T}
\end{equation}
resulting that the susceptibility is an entanglement witness because separable states cannot violate the relation showed in Eq.(\ref{suscepav2}). Thus, we can define an entanglement witness based in a thermodynamical observable:
\begin{equation}\label{susceptwit}
\mathcal{W} = \frac{3\,k_{B}\,T\,\bar{\chi}}{(g\mu_{B})^{2}N\,S} - 1
\end{equation}
where there must be entangled states if $\mathcal{W} < 0$ \cite{2005_NJP_7_258}.

As discussed in Refs.[\onlinecite{2007_PRL_98_110502}] and [\onlinecite{2007_NJP_9_46}], it is possible to use the such entanglement witness to obtain quantitative estimates of a given entanglement measure like negativity\cite{2005_PRA_72_032309}. So, this thermodynamical witness is of great importance because gives the connection between the experimental magnetic susceptibility and the existence of entangled states in a magnetic compound, without needing tomographic knowledge. Also, from Eq.(\ref{susceptwit}) it is possible to determine a temperature of entanglement $T_e$, i.e., the one for which the $\mathcal{W}\rightarrow 0$ and consequently that {\it indicates the entanglement-separability frontier of the system} \cite{2009_EPL_87_40008}.

To obtain $T_e$, let us consider a Hamiltonian of a spin$-S$ dimer ($N=2$) composed of a Heisenberg and Zeeman terms:
\begin{equation}
\mathcal{H}=-J\,\textbf{S}_1\cdot \textbf{S}_2-g\mu_{B}\,B\,(S_{1,z}+S_{2,z})
\end{equation}
where $J$ is the exchange interaction, and $B$ is an external magnetic field. Such Hamiltonian has been described quantitatively in the literature as the Heisenberg-Dirac-Van-Vleck Hamiltonian \cite{livro_kahn,livro_reis}. Note that the Heisenberg term commutes with the Zeeman one, obeing the constraint described in the begining of this section and, consequently, allows us to write the magnetic susceptibility in terms of correlation functions.

The magnetic susceptibility for a $\bar{S}=S_1=S_2=5/2$ dimer and $B\rightarrow 0$ \cite{livro_boca,cpc_182_2011_1169} is given by:
\begin{equation}\label{diiiimero}
\chi_d=\frac{2(g\mu_B)^2}{k_BT}\frac{\mathcal{N}}{Z}
\end{equation}
where
\begin{equation}
\nonumber
\mathcal{N}=e^{-\beta E_1}+5e^{-\beta E_2}+14e^{-\beta E_3}+30e^{-\beta E_4}+55e^{-\beta E_5}
\end{equation}
\begin{equation}
\nonumber
Z=1+3e^{-\beta E_1}+5e^{-\beta E_2}+7e^{-\beta E_3}+9e^{-\beta E_4}+11e^{-\beta E_5}
\end{equation}
and $\beta =1/k_BT$. The eigenvalues of energy $E_i$ are in Table \ref{dimero_tabela}. This expression describes the susceptibility of spin dimers with $\bar{S}=S_1=S_2=5/2$. Interestingly, $chi_d$ can be obtained for any $\bar{S}=S_1=S_2$ value, ranging from $\bar{S}=1/2$ up to $\bar{S}=5/2$, through only minor changes in Eq.(\ref{diiiimero}). For $\bar{S}= 5/2-1/2 = 2$, one should suppress the last terms of the numerator and of the denominator, and similarly for $\bar{S}= 2 - 1/2 = 3/2$, one should suppress again the other last terms, and so on, until $\bar{S}= 1/2$.
\begin{table*}
\begin{center}
\begin{tabular}{cccc}
Label & State $|S,m_S\rangle$  & Dege. ($B=0$)& Energy ($B=0)$\\\hline
E$_5$ & $|5,m_S\rangle$ & 11 & $-15J$\\
E$_4$ & $|4,m_S\rangle$ & 9 & $-10J$\\
E$_3$ & $|3,m_S\rangle$ & 7 & $-6J$ \\
E$_2$ & $|2,m_S\rangle$ & 5 & $-3J$\\
E$_1$ & $|1,m_S\rangle$ & 3 & $-J$\\
E$_0$ & $|0,m_S\rangle$ & 1 & $0$ \\\hline
\end{tabular}
\caption{\label{dimero_tabela}States on the basis $|S,m_S\rangle$, degeneracy for zero magnetic field and energy eigenvalues for a $\bar{S}=S_1=S_2=5/2$ dimer. Note $S$ ranges from $|S_1-S_2|$ up to $|S_1+S_2|$.}
\end{center}
\end{table*}

Substituting Eq.(\ref{diiiimero}) in Eq.(\ref{susceptwit}) and calculating the boundary between separable and entangled states, which means $\mathcal{W} = 0$, we obtain a general form for  $T_e \equiv T_e(J,\bar{S})$:
\begin{equation}\label{euro}
T_e = -\frac{9}{20}(2\bar{S}+1)J
\end{equation}
It should be noted that a physical temperature of entanglement ($T_e > 0$) results only for antiferromagnetic clusters, where the exchange parameter $J$ is negative.

From Eq.(\ref{euro}), the most robust entangled states with the highest possible temperature of entanglement, can be obtained by maximizing both, the spin $\bar{S}$ value and the antiferromagnetic exchange parameter. To maximize $\bar{S}$, a high spin (HS) $d^5$ ion is ideal. For first row transitions-metals the possibilities are either Fe(III) or Mn(II). On the other hand, to obtain a negative exchange parameter $J$, following the Goodenough-Kanamori rules, the antiferromagnetic arrangement is obtained when the angle of the $d$--$p$--$d$ orbitals are close to 180$^\circ$. A search in CCSD database \cite{webcsd} revealed that there are 95 di-iron structures with Fe(III) and an angle Fe--O--Fe of 180$^\circ$. In contrast, only 6 di-manganese structures exhibit Mn(II) and an angle Mn--O--Mn of 180$^\circ$. Keeping these observations in mind, we exploited the syntheses of novel iron complexes using imidazole as an organic ligand. Indeed a novel dinuclear Fe(III) complex was found that meets all of the requirements for a high  entanglement temperature.

\section{Crystal structure}\label{crystalstructure}

\begin{figure*}
\begin{center}
\subfigure[Molecular structure with the labeling scheme adopted. Symmetry code: A:$1-x$, $2-y$, $1-z$.]{
\includegraphics[width=7cm]{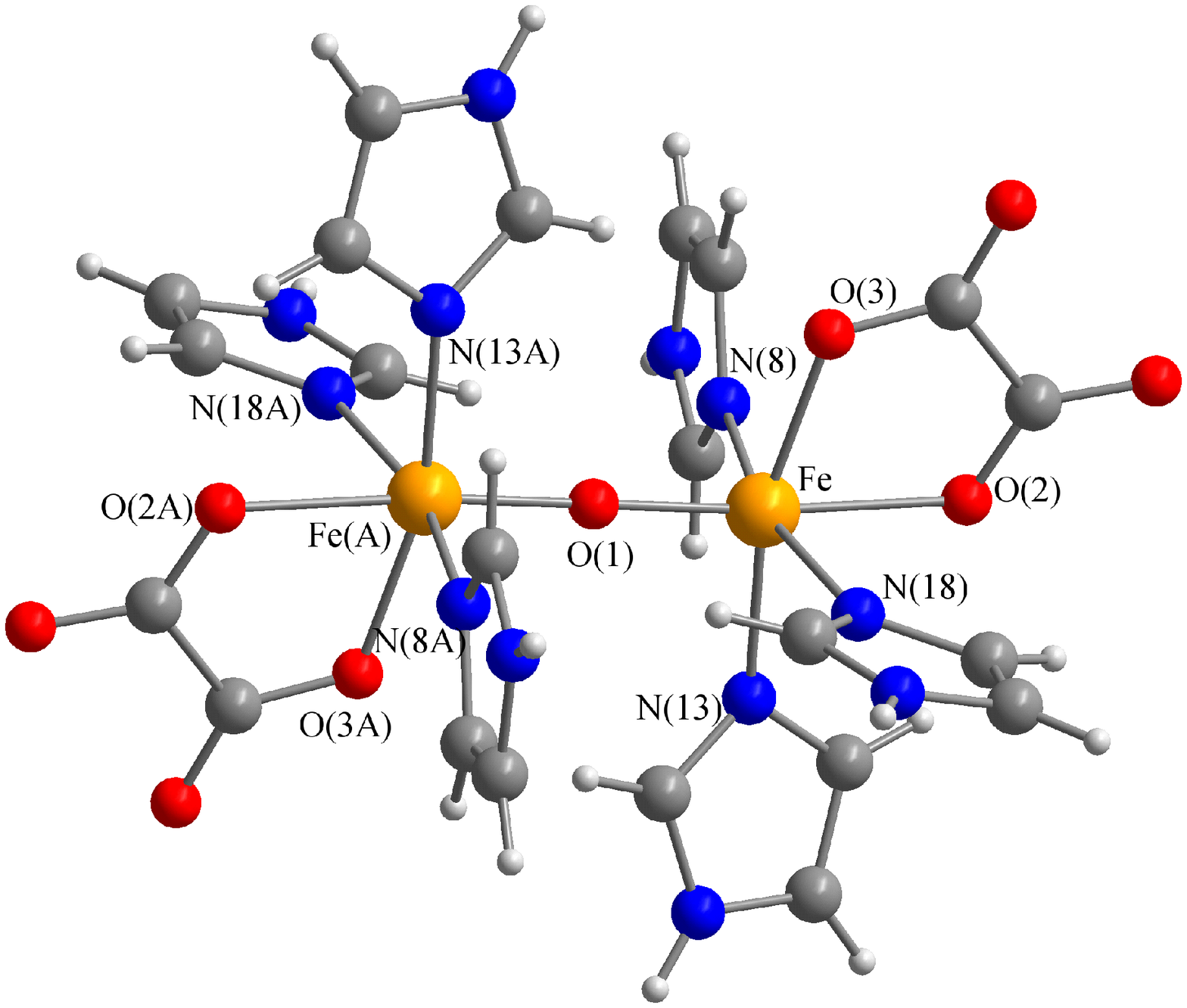}
\label{fig1e}}\hspace{1cm}
\subfigure[Iron polyhedron representation]{
\includegraphics[width=4cm]{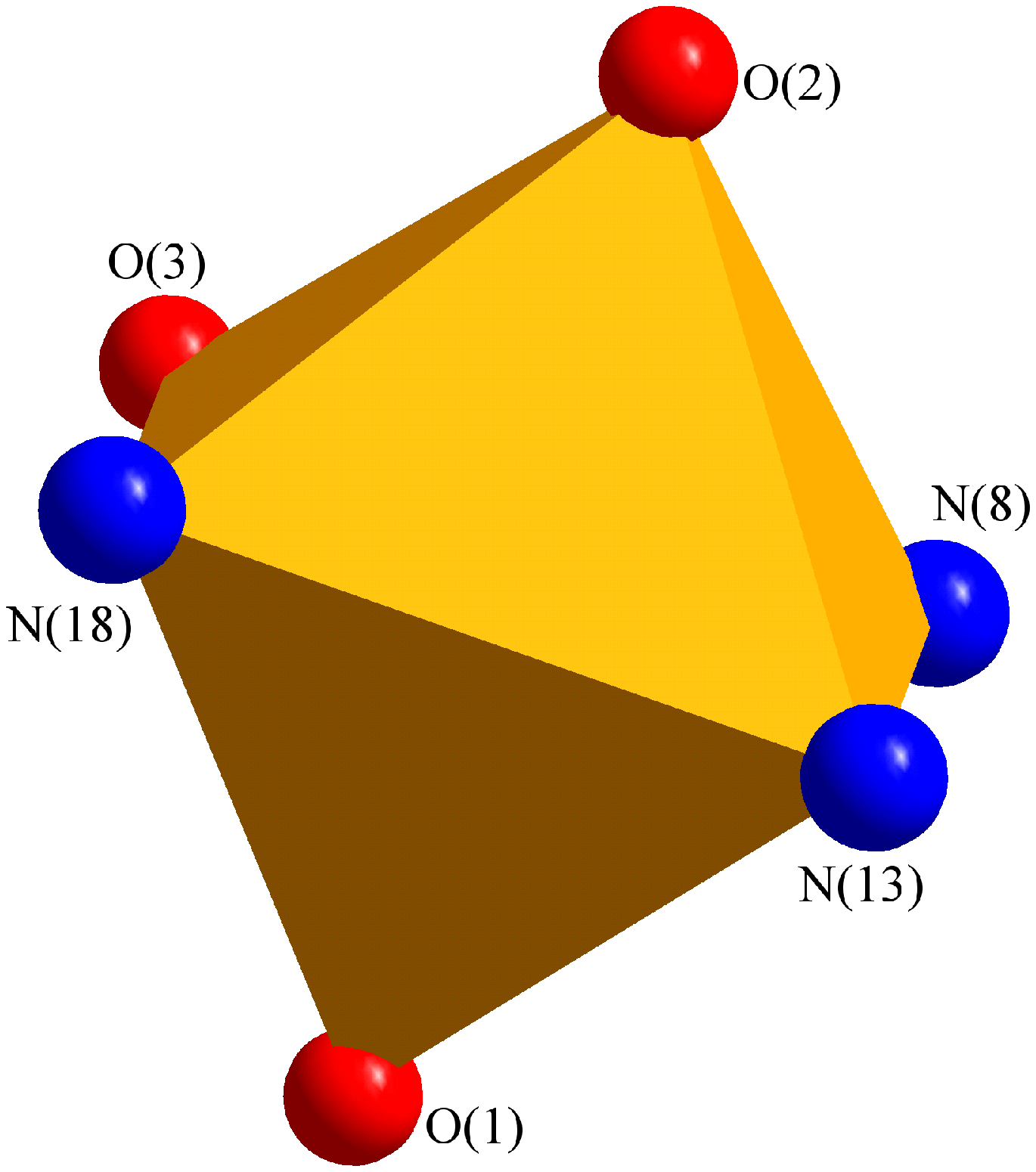}
\label{fig1d}}
\subfigure[Crystal packing diagram along the $b$ axis. The hydrogen bonds are drawn as gray dashed lines.]{
\includegraphics[width=8cm]{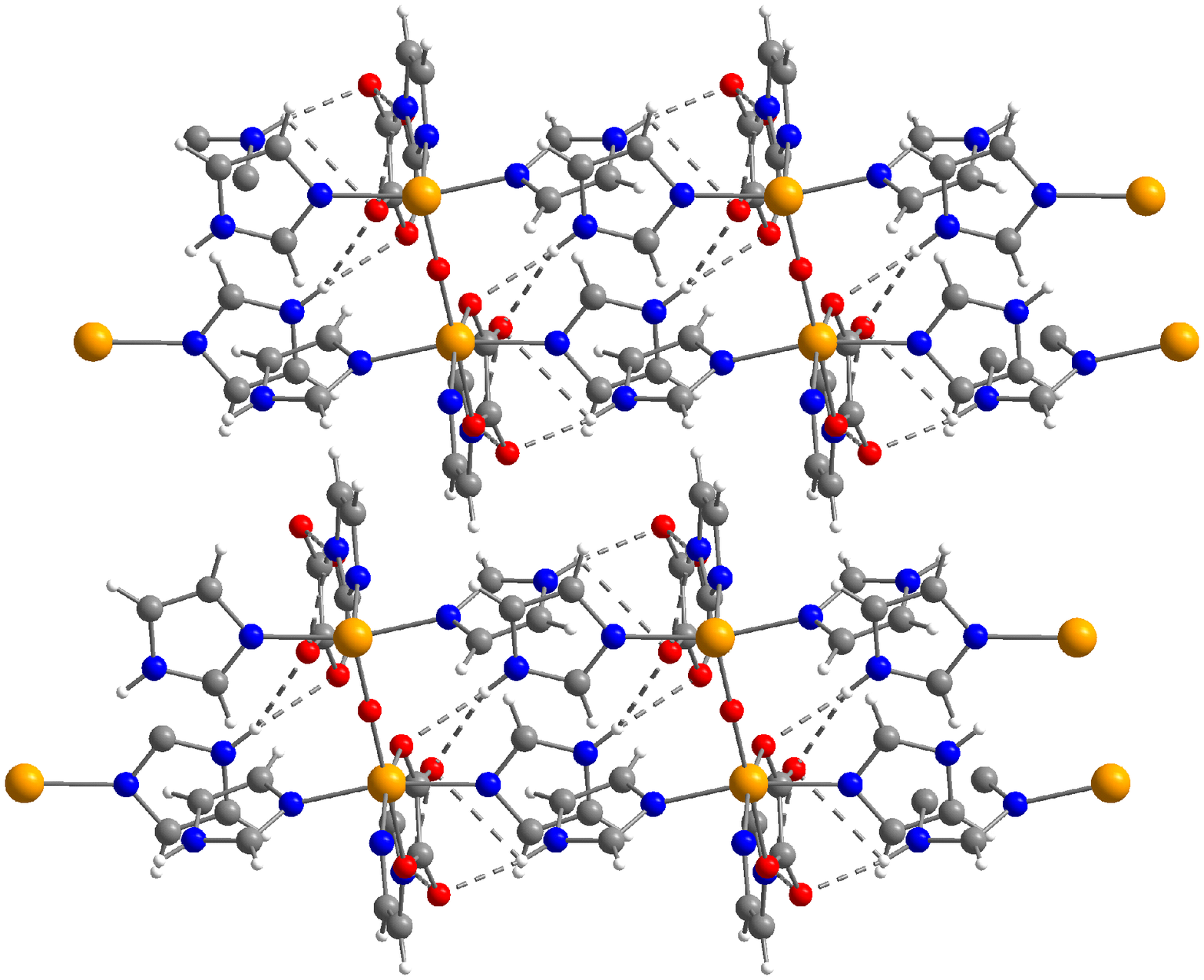}
\label{fig2}}
\caption{Crystal structure details of [Fe$_{2}$($\mu_{2}$-oxo)(C$_{3}$H$_{4}$N$_{2}$)$_{6}$(C$_{2}$O$_{4}$)$_{2}$].}
\end{center}
\end{figure*}

The molecular structure of di-nuclear iron complex is presented in Fig.\ref{fig2}. The crystal structure consists of a di-nuclear $\mu$-oxo Fe(III) complex where the bridging oxygen atom is at the crystallographic inversion center and the metal-to-metal distance is of 3.594(9) {\AA} with a Fe--O--Fe angle of 180$^\circ$. The Fe(III) center is six-coordinated in a distorted octahedra coordination in which the equatorial plane is composed of three nitrogen donors from imidazole ligands (2.095(2), 2.147(2) and 2.168(2) {\AA}), and one oxygen atom of oxalate ligand (2.042(2) {\AA}). The apical positions are occupied by the $\mu$-oxo bridging atom (1.797(5) {\AA}) and by the other oxalate oxygen (2.152(2) {\AA}), as shown in Fig.\ref{fig2}. These Fe--O and Fe--N distances are in the range found for other $\mu$-oxo di-iron(III) complexes\cite{webcsd}. Each di-nuclear iron unit  is linked along the $b$ direction through N--H$\cdots$O hydrogen bonds, between nitrogen atoms from  imidazole ligands and  the oxalate oxygen atoms, with N$\cdots$O distances between 2.858(3) and 3.081(10) {\AA} and N--H$\cdots$O corresponding angles ranging from 122$^\circ$ to 169$^\circ$,  forming one-dimensional (1D) network, effectively isolating the iron dimers (Fig.\ref{fig2}).

\section{Material design}

\begin{figure}
\begin{center}
\includegraphics[width=8cm]{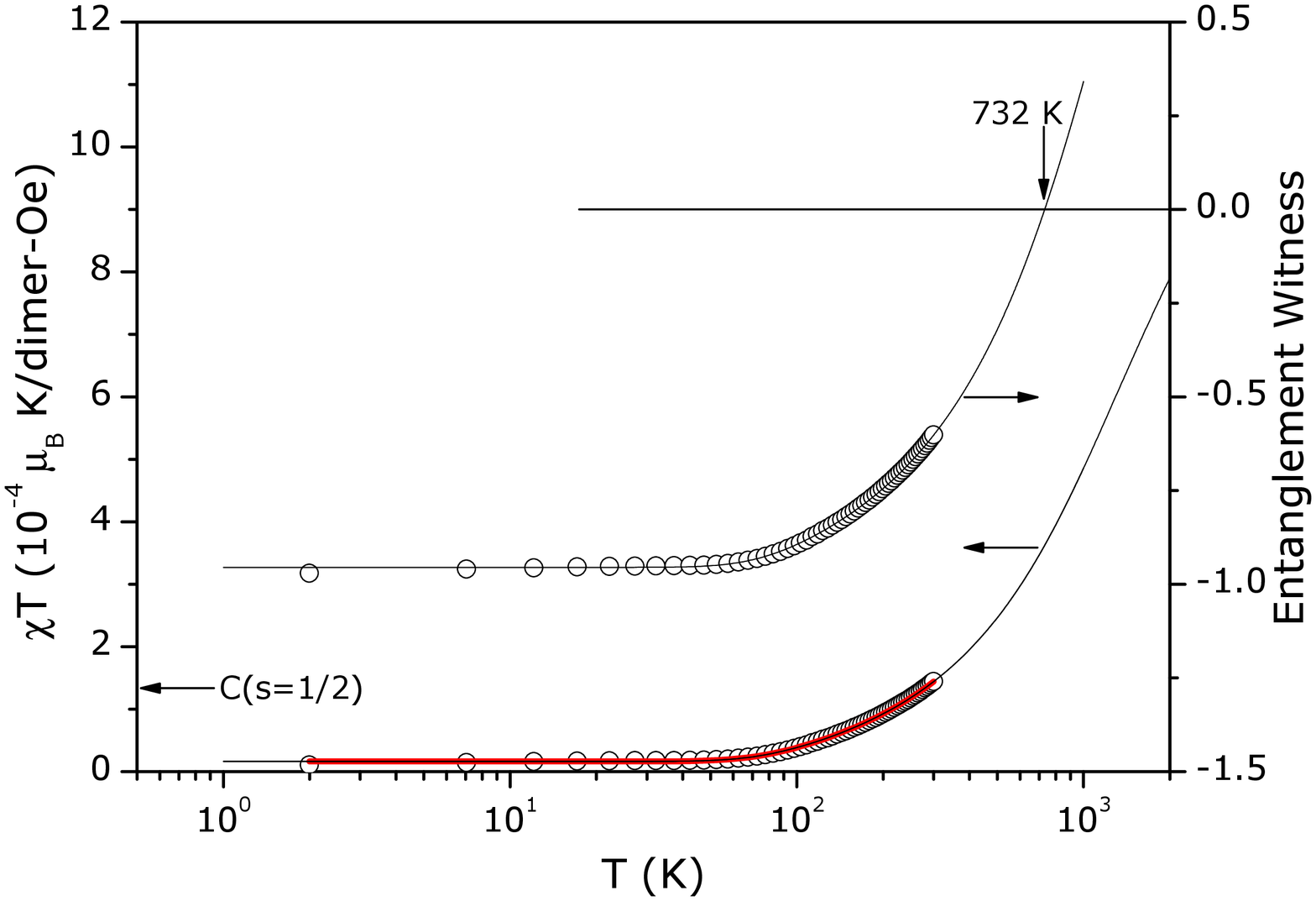}
\caption{Left axis: Experimental (open circles), and theoretical (solid lines) magnetic susceptibility times temperature for the $\bar{S}=S_1=S_2=5/2$ Iron dimer. The thicker red line represents the fitting of Eq.(\ref{vera}) to the data; while the thinner black line is an extrapolation, using the optimized parameters, up to higher temperatures. Right axis: Entanglement witness as a function of temperature. Open circles are from the experimental data, while the solid line is an extrapolation up to higher temperatures using the theoretical model (Eqs.(\ref{susceptwit}) and (\ref{vera})) and the optimized parameter (see text). Note that this material has entangled states up to 732 K.}
\label{fig3}
\end{center}
\end{figure}

As can be seen from Eq.(\ref{susceptwit}), $\mathcal{W}$ is a function of the magnetic susceptibility times temperature, $\chi T$, of the material and, therefore, we measured this quantity to probe $\mathcal{W}$, as presented in Fig.\ref{fig3} (open circles - left axis). The entanglement witness is then easily obtained and is presented also in Fig.\ref{fig3} (open circles - right axis). Up to 300 K, $\mathcal{W}$ remains negative, indicating that the system occupies entangled states up to room temperature. 

As discussed before, to maximize $T_e$ we need to maximize both, $J$ and $\bar{S}$ (see Eq.(\ref{euro})). To verify the HS state of iron in this compound, we can use the magnetic susceptibility. The Curie constant for a free $S=5/2$ ion is $C(S=5/2)=7.84\times 10^{-4}$ $\mu_B$K/Oe, while for a $S=1/2$ free ion, $C(S=1/2)=0.67\times 10^{-4}$ $\mu_B$K/Oe. This last value is indicated by an arrow in Fig.\ref{fig3} multiplied by two, since this figure represents the magnetic susceptibility for the dimer. Using this analysis it can be seen that the Fe(III) is in the HS - $S=5/2$ configuration.

It is noteworthy that the quantity $\chi T$ does not approach zero when the temperature goes to zero (see Fig.\ref{fig3}), this background contribution arises from a small amount $\rho$ of unreacted free Fe(III) ions. This amount does not affect the discussion of the present paper, since our aim is hundreds of Kelvin above room temperature. Taking all these contributions under consideration, the final expression to be fitted to the experimental data is:
\begin{equation}\label{vera}
\chi=(1-\rho)\chi_d + \rho\frac{2C(S=5/2)}{T}
\end{equation}
where $C(S=5/2)=7.84\times 10^{-4}$ $\mu_B$K/Oe is the Curie constant of $S=5/2$ free ion. The red thicker line in Fig.\ref{fig3} is the fit of the above model to the experimental data. The optimized parameters are: $g=1.8$, $J=-282$ K and $\rho=1.0\%$. Note that the first term of the above equation has a factor of 2, because the susceptibility is expressed for a dimer. Based on the results above, it is possible to extrapolate the susceptibility curve up to higher temperatures, as can be seen in Fig.\ref{fig3} (thiner black line). This quantity in Fig.\ref{fig3} tends to the Curie constant of a free $S=5/2$ ion.  

Thus, the $\mathcal{W}$ emerges and we can verify in Fig.\ref{fig3} that this Iron dimer has entangled states that can survive up to 732 K, \textit{hundreds of Kelvin above room temperature}, and well above the expected thermal stability of this compound. This result is in accordance with the empirical prediction of  Eq.(\ref{euro}) which gives  $T_e=761$ K (see Ref.[\onlinecite{2009_EPL_87_40008}]). Therefore this compound shows the highest entanglement temperature reported in the literature, for solid state systems\cite{sciam_vedral}.

\section{Conclusion}

The robust entangled states exhibited by molecular magnets make them among the most promising materials for quantum information processing applications\cite{sciam_vedral}. In order to obtain systems with the highest entanglement temperature, these should contain antiferromagnetically coupled magnetic ions with the highest possible spin states\cite{2009_EPL_87_40008}. A systematic study of molecular iron compounds was made, which resulted in the preparation of a new di-iron complex with formula [Fe$_{2}$($\mu_{2}$-oxo)(C$_{3}$H$_{4}$N$_{2}$)$_{6}$(C$_{2}$O$_{4}$)$_{2}$]. Analysis of magnetic susceptibility data from this compound indicated quantum entanglement (or the entanglement-separability frontier) up to temperatures as high $T_e = 732 K$. This is the highest entanglement temperature reported to date\cite{2009_PRB_79_054408}.

\section{Appendix}\subsection{Synthesis} All reagents and chemicals were purchased from commercial sources and used without further purification. Iron oxalate dihydrate (99\%) and imidazole (99\%) were obtained from Aldrich and Panreac Chemical, respectively.
A mixture of iron oxalate dihydrate (1.39 g; 7.90 mmol), imidazole (0.52 g; 7.90 mmol) and ethanol (70ml), were refluxed for three hours. The resultant brow-orange solution was allowed to cool down overnight. A light brown coloured crystalline product was obtained after 3 weeks.

\subsection{Crystallography} The newly synthesized Fe(III) complex has been structurally characterized by single crystal X-ray crystallography. The complex with formula [Fe$_{2}$($\mu_{2}$-oxo)(C$_{3}$H$_{4}$N$_{2}$)$_{6}$(C$_{2}$O$_{4}$)$_{2}$] crystallizes in the triclinic crystal system with the space group P$\bar{1}$ where a = 8.4789(11), b = 9.6481(12), c = 9.769(2) {\AA}, $\alpha$ = 104.350(7), $\beta$ = 91.673(7), $\gamma$ = 115.215(5) $^\circ$,V = 692.1(2) {\AA}$^3$, Z = 1.
The X-ray data was collected on a CCD Bruker APEX II at 150(2) K using graphite monochromatized Mo-K$\alpha$ radiation ($\lambda$ = 0.71073 {\AA}). The crystal was positioned at 35 mm from the detector and the frames were measured using a counting time of 60 s. A total of 7211 reflections were collected and subsequently merged to 3383 unique reflections with a $R_{int}$ of 0.0374. Data reduction including a multi-scan absorption correction were carried out using the SAINT-NT from Bruker AXS. The structure was solved using SHELXS-97 \cite{actaA_64_2008_112} and refined using full-matrix least squares in SHELXL-97 \cite{actaA_64_2008_112}. The carbon and nitrogen bonded hydrogen atoms were included at calculated positions. Anisotropic thermal parameters were used for all non-hydrogen atoms while the hydrogen atoms were refined with isotropic parameters equivalent to 1.2 times those of the atom to which they were attached. Two atoms N(21) and C(22) from one imidazole group were found to be disordered over two positions, and these were introduced in the structure refinement with adjustable occupancies of $x$ and $1-x$, $x$ being equal to 0.568. The final refinement of 225 parameters converged to final $R$ and $R_w$ indices $R_1$ = 0.0411 and $wR_2$ = 0.0856 for 2495 reflections with $I>2$ (I) and $R_1$ = 0.0683 and $wR_2$ = 0.0954 for all $hkl$ data. The molecular diagrams presented are drawn with graphical package software Diamond. The relevant bond distances and angles are reported in Table \ref{table1}. The detailed hydrogen bond distances and angles are reported in Table \ref{table2}.

%\section{Supplementary material}
The crystallographic data for [Fe$_{2}$($\mu_{2}$-oxo)(C$_{3}$H$_{4}$N$_{2}$)$_{6}$(C$_{2}$O$_{4}$)$_{2}$] complex has been deposited in Cambridge Crystallographic Data Centre (CCDC 879631). The data can be obtained free of charge at www.ccdc.cam.ac.uk or from Cambridge Crystallographic Data Centre, 12 Union Road, Cambridge CB2 1EZ, UK; fax +44 (0) 1223 336033, e-mail: deposit@ccdc.cam.ac.uk.

\begin{table*}
\begin{center}
\begin{tabular}{ll|ll|ll}
\textbf{Bond lenghts (\AA)}\\\hline
Fe--O(1) & 1.797(5) & Fe--O(2)&2.152(2)&Fe--O(3)&2.042(23)\\
Fe--N(8)	&2.147(2)	&Fe--N(13)	&2.095(2)&	Fe--N(18)&	2.168(2)\\\hline
\textbf{Bond angles ($^\circ$)}\\\hline
O(1)--Fe--O(2)&175.26(5)&O(1)--Fe--O(3)&97.41(5)&O(2)--Fe--O(3)&78.12(7)\\
O(1)--Fe--N(8)&96.56(6)&O(1)--Fe--N(13)&97.47(6)&O(1)--Fe--N(18)&98.79(6)\\
O(2)--Fe--N(8)&81.90(8)&O(2)--Fe--N(13)&87.09(7)&O(2)--Fe--N(18)&82.69(8)\\
O(3)--Fe--N(8)&89.09(8)&O(3)--Fe--N(13)&164.80(7)&O(3)--Fe--N(18)&87.55(8)\\
N(8)--Fe--N(13)&92.29(9)&	N(8)--Fe--N(18)&164.59(8)&N(13)--Fe--N(18)&87.12(9)\\
Fe--O(1)--Fe$^A$&180	&&&&\\\hline
(A) $1-x$, $2-y$, $1-z$.\\
\end{tabular}
\caption{Selected bond lengths ({\AA}) and angles ($^\circ$) for [Fe$_{2}$($\mu_{2}$-oxo)(C$_{3}$H$_{4}$N$_{2}$)$_{6}$(C$_{2}$O$_{4}$)$_{2}$].}\label{table1}
\end{center}
\end{table*}

\begin{table*}
\begin{center}
\begin{tabular}{l|c|c|c}\hline
D--H$\cdots$A & H$\cdots$A (\AA) & D$\cdots$A (\AA) & D--H$\cdots$A  (\AA)\\\hline
N(11)--H(11)$\cdots$O(7) [$-1+x$,$y$,$z$]  & 2.05 & 2.922(4) & 169\\
N(16)--H(16)$\cdots$O(5) [$x$,$1+y$,$z$]  & 2.45 & 3.013(3) & 122\\
N(16)--H(16)$\cdots$O(7) [$x$,$1+y$,$z$]  & 2.01 & 2.858(3) & 161\\
N(21)--H(21)$\cdots$O(3) [$2-x$,$2-y$,$1-z$]  & 2.48 & 3.081(10) & 126\\
N(21)-H(21)$\cdots$O(5) [$2-x$,$2-y$,$1-z$]  & 2.17 & 3.028(12) & 164\\\hline
\end{tabular}
\caption{Hydrogen bond dimensions for [Fe$_{2}$($\mu_{2}$-oxo)(C$_{3}$H$_{4}$N$_{2}$)$_{6}$(C$_{2}$O$_{4}$)$_{2}$].}\label{table2}
\end{center}
\end{table*}

\section*{Acknowledgement}

MSR thanks FAPERJ, CAPES, CNPq and PROPPi-UFF for the financial support. AMS (QCMD)  acknowledge support from the Scientific User Facilities Division, and the Office of Basic Energy Sciences of the US Dept. of Energy. BCS was supported by the US Department of Energy, Office of Basic Energy Sciences, Materials Sciences and Engineering Division. DOSP would like to thank the Brazilian funding agency CNPq for the financial support. P. Brand\~{a}o thanks the collaboration project FCT/CAPES/2011/2012 between Portugal and Brasil. The authors would like to thank Dr. P. Hyllus for indicating Ref.[\onlinecite{2009_PhysRep_474_1}] and Prof. G. T{\'o}th for fruitful correspondence.

\end{document}